\documentclass[cameraready]{Interspeech}
\title{Relating the Neural Representations of Vocalized, Mimed, and Imagined Speech}

\author[affiliation={1}]{Maryam}{Maghsoudi}
\author[affiliation={1}]{Rupesh}{Chillale}
\author[affiliation={1}]{Shihab A.}{Shamma}
\address{
    $^1$ University of Maryland - College Park, United States
    }

\email{maryam00@umd.edu, rupesh.chillale@ahduni.edu.in, sas@umd.edu}


\usepackage{comment}
\usepackage{todonotes}

\begin{document}

\maketitle

\begin{abstract}
    We investigated the relationship among neural representations of vocalized, mimed, and imagined speech recorded using publicly available stereotactic EEG recordings. Most prior studies have focused on decoding speech responses within each condition separately. Here, instead, we explore how responses across conditions relate by training linear spectrogram reconstruction models for each condition and evaluate their generalization across conditions. We demonstrate that linear decoders trained on one condition generally transfer successfully to others, implying shared speech representations. This commonality was assessed with stimulus-level discriminability by performing a rank-based analysis demonstrating preservation of stimulus-specific structure in both within- and across-conditions. Finally, we compared linear reconstructions to those from a nonlinear neural network. While both exhibited cross-condition transfer, linear models achieve superior stimulus-level discriminability. 
    % We investigated the relationship among the neural representations of vocalized, mimed, and imagined speech recorded using publicly available stereotactic EEG recordings. Most prior studies have focused on decoding speech responses within each condition separately. Here instead we explore how responses in these different conditions relate to one another by training a linear reconstruction of the speech spectrograms for each condition and evaluating how well they generalize across-conditions. We demonstrate that linear decoders trained on one condition generally transfer successfully to others, implying substantial shared representations of the speech responses. This commonality was assessed with stimulus-level discriminability by performing a rank-based analysis that also demonstrated the preservation of stimulus-specific structure in both within- and across-conditions. Finally, we compared linear reconstructions to those from nonlinear neural network architectures. The results confirmed the transfer across-conditions, but also demonstrated that linear reconstruction models achieve superior stimulus-level discriminability. 
\end{abstract}

\section{Introduction}
Understanding how neural responses to speech reflect its phonetic, syllabic, and semantic features is a central aim of auditory neuroscience, providing insights into both the processing of speech in the brain, and the development of brain–computer interfaces (BCI) \cite{makin2020machine}. Numerous studies have attempted to decode speech and its features from invasive and non-invasive brain responses to external auditory stimuli, i.e., during \textit{listening} to speech. For example, most prior work has reconstructed speech spectrograms from auditory cortical activity and identified phonetic feature encoding during passive listening \cite{mesgarani2014phonetic, pasley2012reconstructing, akbari2019reconstructing}.  But recent reports have demonstrated high-fidelity reconstruction and synthesis of overt speech from neural responses recorded during speech vocalization \cite{anumanchipalli2019speech}, an experimental condition that is important for advances in BCI and for elaborating  the internal processes to produce speech vocalization, such as planning and motor articulation \cite{hickok2007cortical, indefrey2004spatial}. 

In addition to vocalized speech, two additional forms of speech are relevant for understanding internal speech processes: mimed and imagined speech. Mimed speech involves articulatory planning and motor execution without overt acoustic output or auditory feedback. Imagined speech, more accurately imagining to speak, reflects internal generative and planning processes without overt motor movement or acoustic output. Recent work has investigated decoding from mimed \cite{martin2014decoding} and imagined speech \cite{moses2021neuroprosthesis, willett2023high, martin2016word}. These studies demonstrated that silent articulation and attempted speech contain structured neural information sufficient for decoding. These two conditions are even more important in BCI applications, as they address individuals without the ability to produce speech. Note that decoding imagined speech presents additional challenges due to the absence of external timing cues, which complicates alignment between neural activity and speech features \cite{maghsoudi2025convolutional}. 

In this work, in addition to decoding speech from these three conditions, we are specifically interested in neural representation similarities among them. Prior work has reported overlapping neural activity between overt and imagined speech \cite{pei2011spatiotemporal, tian2010mental} motivating investigation of their representational similarity and suggesting that decoders trained on one condition may generalize to others. For instance, a decoder trained within one condition learns a mapping between neural activity patterns and its acoustic features. But when applied to a different condition, its performance would depend on the extent to which the underlying neural representations are shared between the two.

Here we shall use a publicly available stereotactic EEG dataset containing recordings of vocalized, mimed, and imagined speech of a single participant \cite{He2025VocalMind} . We \textit{first} fit linear decoders separately for each condition and evaluate reconstruction performance within each condition. \textit{Second}, we perform cross-condition evaluation to test whether decoders trained in one  case can be applied to the others. In this part, we also evaluate whether there is shared neural representational similarity between the three conditions that allows for such decoder transfer. \textit{Third}, to evaluate how much stimulus-specific information is preserved during speech reconstruction, we perform a rank-based analysis to measure stimulus-specific discriminability in reconstructed speech. While reconstruction correlation provides a measure of overall similarity between predicted and target speech features, it does not directly quantify whether stimulus-specific structure is preserved. For this reason, it is important to complement correlation-based evaluation with analyses that assess stimulus-level individuality. \textit{Fourth}, linear models are known to be more interpretable, which allows for reasoning about possible shared neural representations. Moreover, recent work has shown that linear decoders can perform competitively with nonlinear architectures in reconstructing speech acoustics from neural signals \cite{macintyre2026decoding}. To examine whether these results are consistent across different decoder architectures, we perform the same analyses using a nonlinear decoder introduced in prior work \cite{He2025VocalMind}, which uses convolutional and recurrent layers for spectrogram reconstruction. Therefore, altogether this work presents an investigation of the relationships among neural representations of three speech modes of production (vocalized, mimed, and imagined) that may further help with a deeper understanding of the process, and eventually with BCI applications.

\section{Methods}
\subsection{Dataset}
We used the publicly available VocalMind dataset \cite{He2025VocalMind}, which includes stereotactic EEG (sEEG) recordings from a single participant during vocalized, mimed, and imagined speech in Mandarin Chinese. The dataset contains 100 sentences, each repeated twice per condition. We used the published preprocessed signals from 110 electrodes (low-pass filtered at 100 Hz). For stimulus representation, we used the provided vocalized audio recordings and computed time–frequency features using the NSL cortical model \cite{Chi2005Multiresolution} which mimics the processing of the human cochlea by transforming the acoustic waveform into a biologically motivated time–frequency representation.

\subsection{Linear reconstruction model}
We trained linear decoders to reconstruct the stimulus spectrogram from time-lagged sEEG responses. The reconstructed spectrogram value at time $t$ and frequency $f$ is given by:
\begin{equation}
    \hat{S}(t,f) = \sum_{c} \sum_{l \in L} g(c,l,f)\, X(t-l,c)
\end{equation}
where $g(c,l,f)$ are the model weights for each channel $c$, lag $l$, and frequency $f$, and $X(t-l,c)$ is the neural response at channel $c$ lagged by $l$. $L$ represents the set of all lags used.

In matrix form, the above equation can be written as $\hat{S} = G^{T} X_{\text{lag}}$, and the solution is given by:
\begin{equation}
    G = \arg\min_{G} || S - G^{T} X_{\text{lag}} ||_F^2 + \alpha ||G||_F^2
\end{equation}
The regularization parameter $\alpha$ was selected using grid search to prevent overfitting. We split the dataset into training and test sets and fit $G$ separtely for each condition (Vocalized, Mimed, and Imagined), yielding transfer functions called $G_v$, $G_m$, and $G_i$. 

To assess statistical significance of these models, we fit null models by shuffling the trial orders, thereby pairing sEEG responses with mismatched spectrograms, resulting in $G_{v-null}$, $G_{m-null}$, and $G_{i-null}$. All other computations were kept the same.

\subsection{Non-linear reconstruction model}
\begin{figure}[t]
  \centering
  \includegraphics[width=\linewidth]{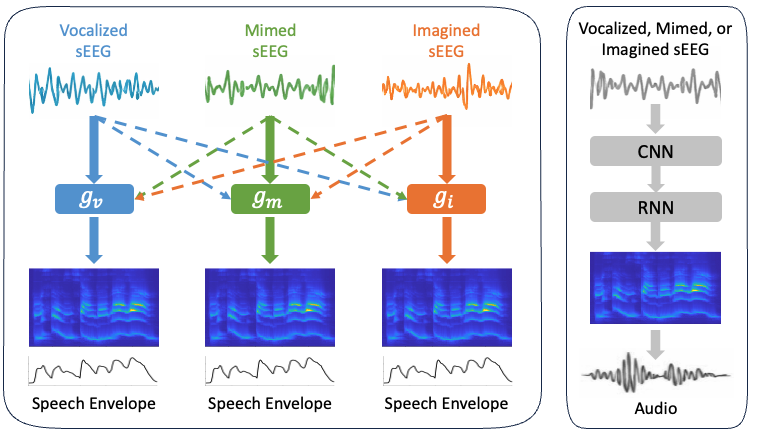}
  \caption{Linear and nonlinear decoding approaches. Left: Condition-specific linear decoders are trained on Vocalized, Mimed, or Imagined sEEG and evaluated both within and across conditions. Right: Nonlinear neural network architecture used for spectrogram reconstruction in \cite{He2025VocalMind}.}
  \label{fig:main}
\end{figure}

We retrained the model introduced in the VocalMind study \cite{He2025VocalMind} to decode mel-spectrograms from sEEG signals. The architecture includes a convolutional block for feature extraction followed by a recurrent (RNN) block that captures temporal dependencies (Figure \ref{fig:main}, left). Mean squared error (MSE) loss is used during training. The estimated mel-spectrogram can be converted into speech using a pre-trained HiFi-GAN model \cite{kong2020hifigan}. 

Training was performed separately for Vocalized, Mimed, and 
Imagined responses, producing reconstructed spectrograms for each condition, from which envelopes are computed. These reconstructions are evaluated using the same metrics described in the following section, allowing comparison between linear and nonlinear approaches.

\subsection{Evaluation}
We evaluate cross-condition generalization to assess how well 
decoders transfer across speaking conditions. In cross-condition evaluation, the model is trained on one condition and tested on the held-out data from the same condition as well as the other two conditions (Figure \ref{fig:main}, left). To measure the similarity between the reconstructed and the target spectrograms, we compute the linear correlation between them. Additionally, we compute the spectrogram envelope by averaging 
over the frequency axis and measure the linear correlation 
between reconstructed and target envelopes. We compare these correlation values with those obtained from null models and measure statistical significance.

To see how well the sentence-level discriminability is preserved in the reconstructed spectrograms, we perform rank analysis. For each reconstructed spectrogram, the envelope is computed and is correlated with all candidate target envelopes. For each $k$ we ask the question: In how many cases does the correct sentence appear within the top-k matches? We plot the top-k curve and compare it to chance, where $chance = \frac{k}{N}$ and $N$ is the total number of sentences. An early large gap between the top-k curve and chance shows stronger sentence-specificity. We evaluate the significance by computing the difference in the area under curve (AUC) of the top-k curve and the chance line. 
\begin{equation}
    \mathrm{AUC} = \sum_{k=1}^{N} (\mathrm{Top}_k(k) - \frac{k}{N})
\end{equation}
Higher AUC values indicate stronger sentence-level discriminability in the reconstructed signals.

\section{Results}

\subsection{Cross-condition reconstruction performance}

\begin{figure}[t]
  \centering
  \includegraphics[width=\linewidth]{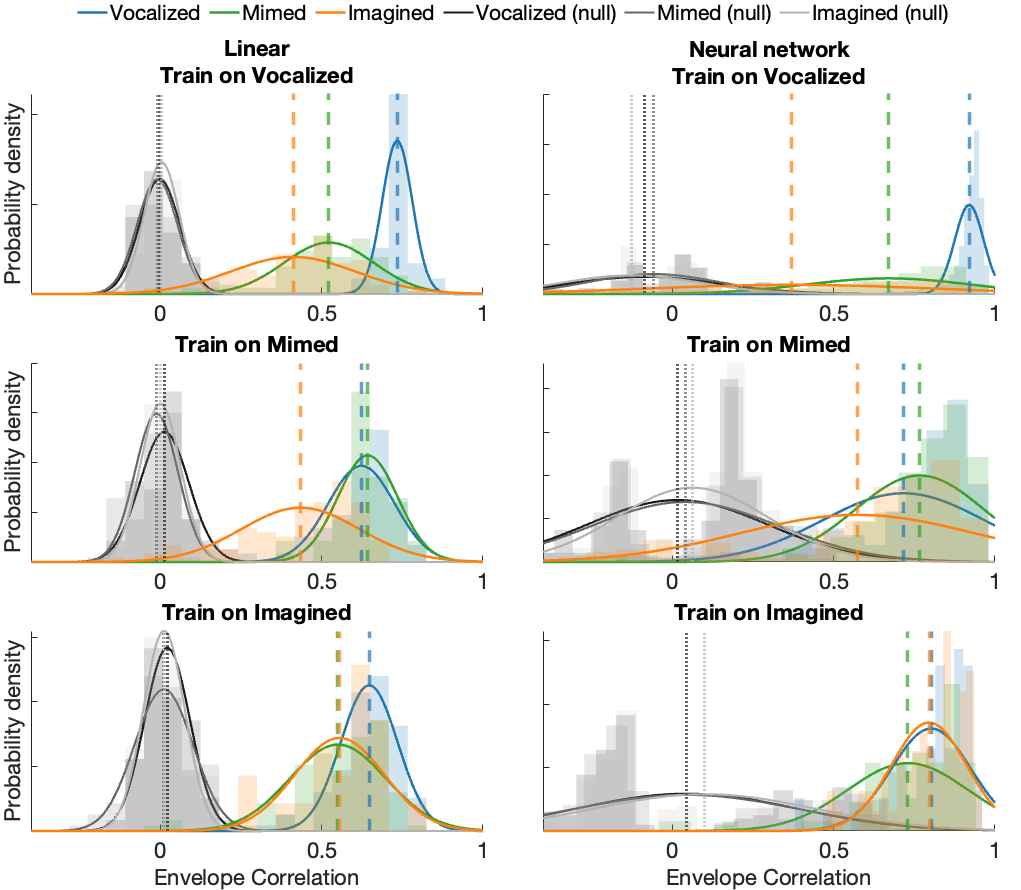}
  \caption{Envelope correlation distributions for linear (left) and neural network (right) decoders. Rows indicate training condition. Colored curves show performance across test conditions; gray distributions represent null models obtained by shuffling. Dashed lines mark mean correlations.}
  \label{fig:res1}
\end{figure}

We evaluate speech envelope reconstruction within and across conditions. 
In the within-condition setting, $G_{\text{condition}}$ is applied to 
held-out test sentences from the same condition. In the cross-condition 
setting, $G_{\text{condition}}$ is applied to the other two conditions 
(e.g., $G_m$ applied to $X_v$ and $X_i$). Reconstruction performance 
is quantified using linear correlations between reconstructed and target 
spectrogram envelopes. In all cases, correlation distributions significantly 
exceed their corresponding null distributions (all $p \ll 0.001$), generated 
using $G_{v\text{-null}}$, $G_{m\text{-null}}$, and $G_{i\text{-null}}$.

Figure~\ref{fig:res1} (left) summarizes cross-condition performance for 
linear models. The Vocalized-trained decoder ($G_v$) achieves the highest 
performance on Vocalized data ($p \ll 0.001$), followed by Mimed and then 
Imagined data, indicating progressively reduced transfer as articulatory 
and sensory components are removed.

For the Mimed-trained decoder ($G_m$), performance on Mimed and Vocalized 
data does not differ significantly ($p = 0.6$), while both exceed performance 
on Imagined speech ($p \ll 0.001$). This indicates strong shared 
articulatory-related structure between Mimed and Vocalized conditions, 
with reduced overlap for Imagined speech.

For the Imagined-trained decoder ($G_i$), performance is highest on Vocalized data ($p \ll 0.005$) and comparable between Imagined and Mimed conditions. Although this ordering may appear counterintuitive, it indicates that 
the additional neural processes present during Vocalized speech do not obscure the planning-related information learned from Imagined speech, and may even preserve it.

Overall, Mimed-trained decoders treat 
Mimed and Vocalized similarly, while Imagined-trained decoders treat 
Imagined and Mimed similarly. This supports the view that Mimed speech 
occupies an intermediate representational position between Vocalized 
and Imagined speech, sharing articulatory-related structure with the 
former and planning-related structure with the latter.

\subsection{A linear explanation for cross-condition transfer}
Let $X_{\text{lag}}$ denote the neural response matrix used for decoding. 
We assume neural activity can be decomposed into components reflecting 
different processing stages: Imagined speech primarily contains internal 
planning or generative processes ($X_P$), Mimed speech contains planning 
plus articulatory activity ($X_P + X_A$), and Vocalized speech contains 
planning, articulatory, and sensory/feedback-related components 
($X_P + X_A + X_S$). Formally,
\begin{equation}
X_V = X_P + X_A + X_S, \qquad
X_M = X_P + X_A, \qquad
X_I = X_P.
\end{equation}
The linear decoder is
$\hat{S} = G^{T} X_{\text{lag}}$,
where $G$ is learned from a single training condition. Because 
$G_{\text{train}}$ is optimized only on that condition, it captures 
neural directions that are predictive within that training regime. 
Geometrically, the decoder defines a neural subspace used to reconstruct 
the stimulus, and cross-condition generalization depends on how much 
test-condition activity projects onto this learned subspace (i.e., subspace overlap).

For a Mimed-trained decoder (trained on $X_M = X_P + X_A$), applying it to
Vocalized responses gives
\begin{align}
\hat{S}_{M\rightarrow V} 
&= G_M^{T} X_V 
 = G_M^{T}(X_P + X_A + X_S) \notag\\
&= \underbrace{G_M^{T}(X_P + X_A)}_{\hat{S}_{M\rightarrow M}}
 + G_M^{T}X_S,
\end{align}
so, $\hat{S}_{M\rightarrow V} - \hat{S}_{M\rightarrow M} = G_M^{T}X_S$.
Thus, $M\rightarrow V$ is similar to $M\rightarrow M$ when the additional
Vocalized-specific component $X_S$ has small projection onto the predictive
subspace captured by $G_M$. In contrast, applying the same decoder to Imagined
speech yields
\begin{equation}
\hat{S}_{M\rightarrow I} = G_M^{T}X_I = G_M^{T}X_P,
\end{equation}
and relative to Mimed testing,
$\hat{S}_{M\rightarrow M} - \hat{S}_{M\rightarrow I} = G_M^{T}X_A,$
predicting $M \rightarrow M \approx M \rightarrow V > M \rightarrow I$ when
$X_A$ contributes to stimulus prediction.

For an Imagined-trained decoder (trained only on $X_I = X_P$), additional 
components present in Mimed and Vocalized speech ($X_A$, $X_S$) are not 
explicitly learned during training, but they can still influence 
reconstruction if they overlap with the planning subspace. To make this 
precise, decompose $X_A = X_A^{\parallel} + X_A^{\perp}$ and 
$X_S = X_S^{\parallel} + X_S^{\perp}$, where the $\parallel$ components 
lie in the imagined/planning subspace and the $\perp$ components are 
orthogonal to it. Because $G_I$ is learned from $X_P$, it primarily acts 
on directions within this subspace, implying 
$G_I^{T}X_A = G_I^{T}X_A^{\parallel}$ and 
$G_I^{T}X_S = G_I^{T}X_S^{\parallel}$. 

Thus, articulatory and sensory activity contribute to reconstruction only 
to the extent that they project onto planning-related directions. If these 
projections are substantial, the Imagined-trained decoder can generalize 
to Mimed (and potentially Vocalized) speech; if not, performance drops. 

\subsection{Sentence-level discriminability}

While envelope correlation captures overall reconstruction 
similarity, it does not directly measure preservation of 
sentence-specific structure. To evaluate discriminability, we 
perform rank analysis between reconstructed and target envelopes 
and compute the AUC above chance for the top-$k$ curves for 
each trained model ($G_v$, $G_m$, $G_i$) applied to each condition 
($X_v$, $X_m$, $X_i$).

Sentence-level discriminability is strongest when models are 
tested on Vocalized responses. The highest AUC is obtained 
for $G_v$ applied to $X_v$ (AUC = 0.32). When $G_m$ and 
$G_i$ are applied to $X_v$, performance remains above chance, 
indicating preserved sentence-specific structure in Vocalized data.

For the Mimed-trained decoder, performance on $X_v$ and 
$X_m$ is similar (AUC = 0.11 and 0.13) and substantially 
lower on $X_i$ (AUC = -0.01). 

For the Imagined-trained decoder, performance on $X_i$ and 
$X_m$ is comparable (AUC = 0.07 and 0.08), and lower overall 
relative to Vocalized testing.

Overall, the rank analysis shows the same transfer 
structure observed in reconstruction correlations: Mimed-trained 
decoders treat Mimed and Vocalized similarly, while Imagined-trained 
decoders treat Imagined and Mimed similarly. This supports the view 
that Mimed speech occupies an intermediate representational position 
between Vocalized and Imagined speech.

\begin{figure}[t]
  \centering
  \includegraphics[width=220pt]{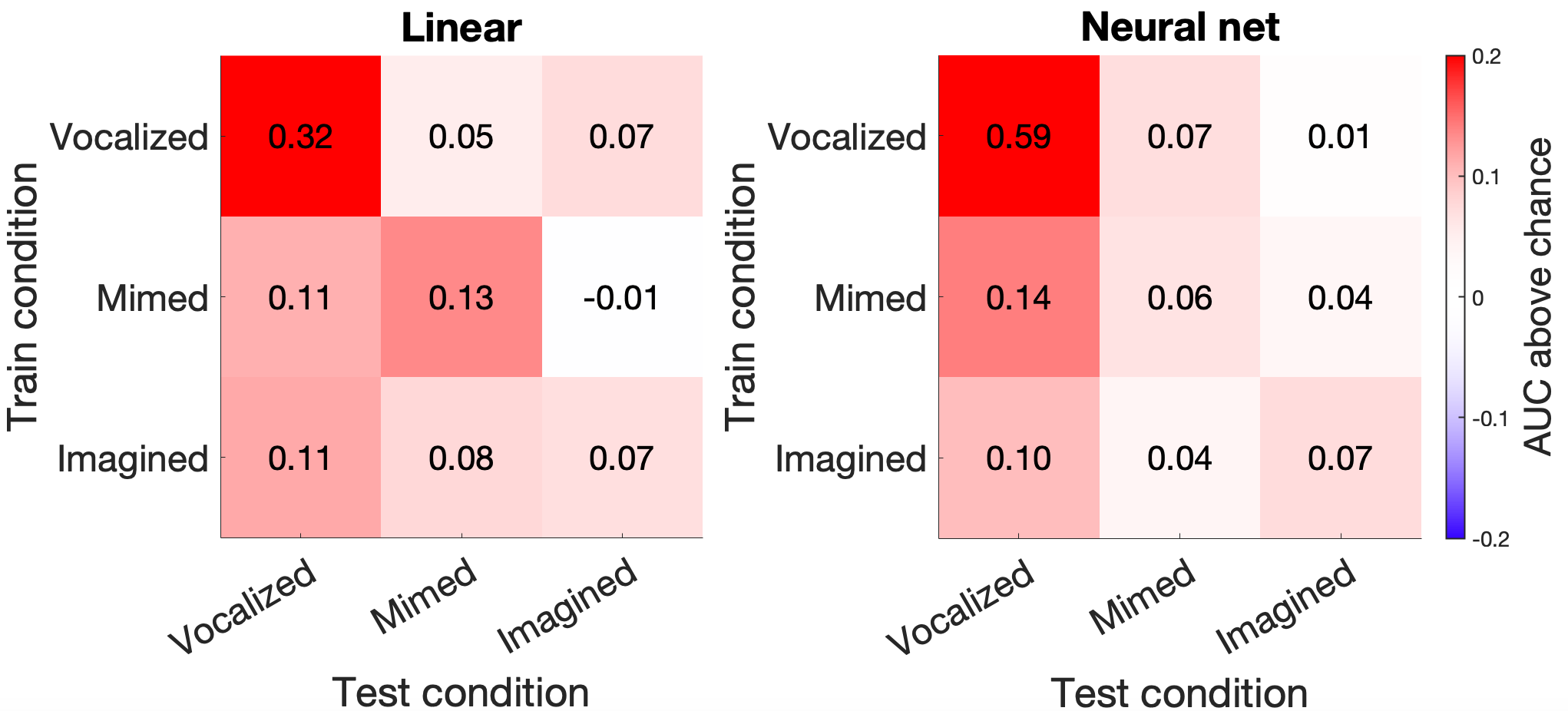}
  \caption{Sentence-level discriminability measured by AUC above chance for linear (left) and neural network (right) decoders. Positive values (red) reflect performance above chance.}
  \label{fig:auc}
\end{figure}

\subsection{Non-linear neural network decoder}
We next evaluate cross-condition reconstruction performance 
using the nonlinear neural network model. Figure~\ref{fig:res1} 
(right) shows reconstructed envelope correlations for each 
training–test pairing. Null models were trained with temporally circularly shifted target spectrograms. The nonlinear model significantly 
outperforms null in both within- and cross-condition settings 
(all $p \ll 0.001$), indicating reliable transfer across conditions.

When trained on Vocalized speech, the model achieves its 
highest performance on Vocalized responses and remains 
significantly above null for Mimed and Imagined speech 
($p \ll 0.001$), though reduced relative to the within-condition 
case.

For the Mimed-trained model, performance on Mimed and 
Vocalized responses remains close ($p = 0.02$, $d = 0.16$), 
with both exceeding performance on Imagined speech, 
consistent with shared structure between Vocalized and Mimed activity.

For the Imagined-trained model, performance is comparable 
between Vocalized and Imagined responses ($p = 0.74$, 
$d = 0.04$) and lower on Mimed speech ($p = 0.02$, $d = 0.29$). Overall, the nonlinear model shows a similar transfer pattern to the linear model, with a difference in the Imagined-trained case.

To assess sentence-level discriminability, we apply the same 
rank analysis used in the linear framework. Figure~\ref{fig:auc} 
(right) shows AUC above chance for the nonlinear model. 
Training and testing on Vocalized speech yields the highest 
AUC (0.59), and testing on Vocalized responses consistently 
produces the strongest discriminability across all training 
conditions. For the Mimed-trained network, performance is highest on 
Vocalized responses (AUC = 0.14), followed by Mimed 
(AUC = 0.06), and lowest on Imagined speech (AUC = 0.04). 
Similarly, for the Imagined-trained network, performance 
is highest on Vocalized responses (AUC = 0.10), with lower 
and more comparable values on Imagined and Mimed speech 
(AUC = 0.07 and 0.04).

\begin{figure}[t]
  \centering
  \includegraphics[width=\linewidth]{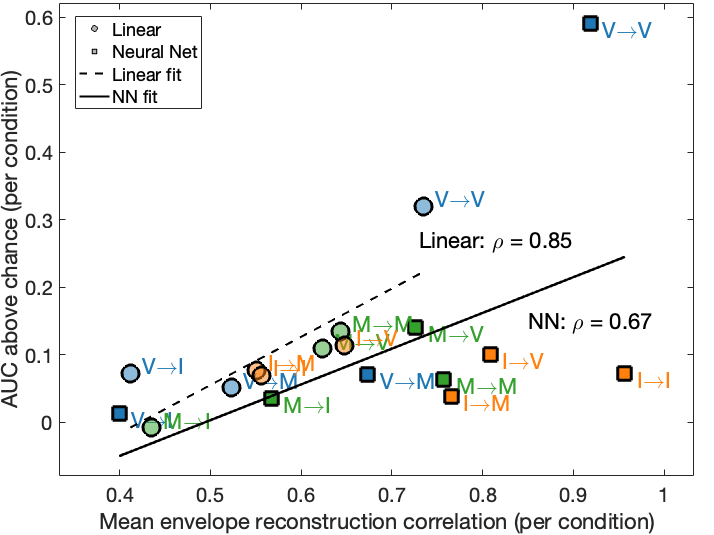}
  \caption{Relationship between mean envelope reconstruction correlation and sentence-level discriminability for linear and neural network decoders. Each point represents a training–test pair (circles: linear; squares: neural network). Dashed and solid lines show the corresponding linear fits.}
  \label{fig:compare}
\end{figure}

Overall, although the nonlinear model achieves higher reconstruction 
correlations ($p \ll 0.005$), its rank analysis differs from the pattern 
observed in the linear model. This difference motivated us to 
examine whether higher correlation necessarily implies stronger 
sentence-level discriminability. As shown in 
Figure~\ref{fig:compare}, AUC increases with mean reconstruction 
correlation for both models; however, the linear model exhibits 
a steeper relationship (Steiger's test, $p = 1.24 \times 10^{-4}$). Improvements in reconstruction 
accuracy therefore do not translate into proportional 
gains in sentence-level separation, and in this setting, the linear model 
better preserves sentence-specific structure.

\section{Discussion}
Our results are consistent with the hierarchical relationship hypothesis between neural responses to Imagined, Mimed, and Vocalized speech, with imagined containing planning-related activity, mimed containing planning and articulatory components, and vocalized containing planning, articulatory, and auditory feedback components. Future research could use subspace analyses, such as comparing principal components or computing subspace alignment between speech conditions, to directly test this hypothesis. Additionally, future work would benefit from incorporating neural recordings during listened speech to evaluate the auditory sensory feedback component hypothesized to be present during vocalization. Finally, our comparison of linear and nonlinear models shows the importance of interpretability when using nonlinear decoders. Future work can further investigate this to better understand which neural components support cross-condition generalization.

\section{Conclusion}
We demonstrated that linear decoders can map sEEG responses to acoustic spectrograms during vocalized, mimed, and imagined speech. When applied cross-conditionally, these decoders still perform significantly above chance, suggesting that neural responses across speech modes share a common structure. We observed transfer patterns consistent with a hierarchical representational organization, where imagined speech engages a subset of processes present in mimed speech, and mimed speech engages a subset of those present in vocalized speech. Rank analysis also showed that stimulus-specific discriminability is preserved above chance in the reconstructions. We also showed that a nonlinear neural network achieves similar cross-condition transfer and above-chance discriminability. However, comparing the linear and nonlinear models demonstrated stronger preservation of stimulus-specific structure in the linear case, emphasizing the importance of interpretability when evaluating neural decoding performance.

\section{Acknowledgement}
This work is supported in part by a grant from the Airforce Office of Scientific Research and NIH R01-DC005779 to SAS.

% Authors must proofread their PDF file prior to submission, to ensure it is correct. Do not rely on proofreading the \LaTeX\xspace source or Word document. \textbf{Please proofread the PDF file before it is submitted.}

% \section{Acknowledgments}

% {\color{blue}Acknowledgments should be included only in the camera-ready version, not in the version submitted for review. For regular papers, pages 5 and 6, and for long papers, pages 9 and 10, are reserved exclusively for acknowledgments, disclosures of the use of generative AI tools, and references. No other content may appear on these pages. Any appendices must be contained within the first four pages for regular papers and within the first eight pages for long papers.

% Acknowledgments and references may begin on an earlier page if space permits.}

% \ifcameraready
%      The Interspeech 2026 organizers
% \else
%      The authors
% \fi
% would like to thank ISCA and the organizing committees of past Interspeech conferences for their help and for kindly providing the previous version of this template.

\section{Generative AI Use Disclosure}
Generative AI tools were used for language editing and minor phrasing refinement. All scientific content and conclusions were developed by the authors.

% \section{Reference Format}

% {\color{blue}It is ISCA policy that papers submitted to Interspeech should refer to peer-reviewed publications. References to non-peer-reviewed publications (including public repositories such as arXiv, Preprints, and HAL, software, and personal communications) should only be made if there is no peer-reviewed publication available, and should be kept to a minimum.

% References should be in standard IEEE format (follows the IEEEtran format), numbered in order of appearance, for example \cite{Davis80-COP} is cited before \cite{Rabiner89-ATO}. For longer works such as books, provide a single entry for the complete work in the References, then cite specific pages \cite[pp.\ 417--422]{Hastie09-TEO} or a chapter \cite[Chapter 2]{Hastie09-TEO}. Multiple references may be cited in a list \cite{Smith22-XXX, Jones22-XXX}.}

\bibliographystyle{IEEEtran}
\bibliography{mybib}

\end{document}